\documentclass[12pt,preprint]{emulateapj}

\usepackage{natbib}
\usepackage{amsmath,amssymb,graphicx,soul,color}

\shorttitle{Faint high-energy gamma-ray photon emission of GRB 081006A from $Fermi$ observations}

\slugcomment{Draft version on November 14 2011}

\begin{document}

\title{Faint high-energy gamma-ray photon emission of GRB 081006A from $Fermi$ observations}

\author{ WeiKang~Zheng\altaffilmark{1}
\email{zwk@umich.edu}
Carl~W.~Akerlof\altaffilmark{1}, 
Shashi~B.~Pandey\altaffilmark{1,2}, 
Timothy~A.~McKay\altaffilmark{1},
BinBin~Zhang\altaffilmark{3,4}, and
Bing~Zhang\altaffilmark{3}
}

\altaffiltext{1} {Department of Physics, University of Michigan, 450 Church Street, Ann Arbor, MI, 48109, USA}
\altaffiltext{2} {Aryabhatta Research Institute of Observational Sciences, Manora Peak, Nainital, India, 263129}
\altaffiltext{3} {Department of Physics and Astronomy, University of Nevada, Las Vegas, NV, 89154, USA}
\altaffiltext{4} {Department of Astronomy \& Astrophysics, Pennsylvania State University, University Park, PA 16802, USA}

\shortauthors{Zheng et al. 2010}

\begin{abstract}
Since the launch of the $Fermi$ $\gamma$ - ray Space Telescope on June 11, 2008, the LAT instrument has solidly detected more than 20 GRBs with high energy photon emission above 100 MeV. Using the matched filter technique, 3 more GRBs have also shown evidence of correlation with high energy photon emission as demonstrated by Akerlof et al. In this paper, we present another GRB unambiguously detected by the matched filter technique, GRB 081006A. This event is associated with more than 13 high energy photons above 100 MeV. The likelihood analysis code provided by the $Fermi$ Science Support Center (FSSC) generated an independent verification of this detection by comparison of the Test Statistics (TS) value with similar calculations for random LAT data fields. We have performed detailed temporal and spectral analysis of photons from 8 keV up to 0.8 GeV from the GBM and the LAT.
The properties of GRB 081006A can be compared to the other two long duration GRBs detected at similar significance, GRB 080825C and GRB 090217A. We find that GRB 081006A is more similar to GRB 080825C with comparable appearances of late high energy photon emission. As demonstrated previously, there appears to be a surprising dearth of faint LAT GRBs, with only one additional GRB identified in a sample of 74 GRBs. In this unique period when both $Swift$ and $Fermi$ are operational, there is some urgency to explore this aspect of GRBs as fully as possible.
\end{abstract}


\section{Introduction}
Gamma-Ray Bursts (GRBs) are extremely luminous explosions in the Universe but most observations are limited to photons
below a few MeV. High energy emission above 100 MeV was
only detected a few times by the EGRET instrument (Dingus et al. 1995) and now more recently 
by AGILE (Giuliani et al. 2008). The recently launched $Fermi$ $\gamma$ - ray Space Telescope has enlarged
the opportunity to study high energy radiation from GRBs by providing large apertures and wide fields of view. Two onboard instruments, the Gamma-ray Burst Monitor (GBM; Meegan et al. 2009) and the Large Area Telescope (LAT; Atwood et al. 2009), overlap
energy bands to span from 8 KeV to above 100 GeV.  With other satellites providing more precise localization ability (e.g. Swift; Gehrels et al. 2004, INTEGRAL; Winkler et al. 2003, AGILE; Giuliani et al. 2008), we have a
unique opportunity to study the physical mechanisms of GRBs across a very wide dynamic energy range.

The $Fermi$/LAT covers the energy range from below 20 MeV to more than 300 GeV with an effective field-of-view (FOV) $\sim$ 2.4 sr (Atwood et al. 2009). The other $Fermi$ instrument, the GBM, is sensitive to the 8 keV - 40 MeV range and covers the entire unocculted sky. The GBM identifies GRBs in real time (Meegan et al. 2009) with a rate of $\sim$ 250 events per year (Paciesas et al. 2010).  During its first 27 months operation, $\sim$150 GRBs have been detected by the GBM at locations that were simultaneously within the $Fermi$/LAT field of view. However, of these simultaneously observed events\footnote{http://fermi.gsfc.nasa.gov/ssc/observations/types/grbs/grb$\_$table/}, only 20 have been detected at a threshold of more than $\sim$ 10 high energy photons above 100 MeV (Granot 2010), corresponding to a rate of $\sim$ 9 GRBs per year.

Using the matched filter technique, the $Fermi$ detection threshold was reduced to a level of $\sim$ 6 high energy photons
with the concomitant identification of 3 additional detections as reported by Akerlof et al. (2010, 2011, hereafter A10, A11), namely GRB 080905A, 091208B and GRB 090228A. In Section 2, we present the first high energy photon detection of GRB 081006A using this matched filter technique. The detailed observations obtained by the GBM and LAT, including temporal and spectral properties, are described in Section 3. The comparison of GRB 081006A with other two GRBs detected with similar fluences are discussed in Section 4 and summarized in Section 5.

\section{Detection of GRB 081006A using the matched filter technique}
Following the method developed by Akerlof et al. in A10 and A11, we apply the matched filter technique to an extended GBM sample by lowering the GBM fluence threshold below the 5 $\mu$erg/cm$^2$ minimum that defined the GRBs listed in Table 1 of A10.
The extended GBM sample consists of either GRBs that have GBM fluences less than 5 $\mu$erg/cm$^2$ or have no GBM fluence information
whatsoever. The remaining criteria are the same as in A10. This new sample of 74 GRBs, listed in Table 1, was searched for high energy photon emission using the matched filter technique. One GRB, namely GRB 081006A, was revealed with a large matched weight value, ${\zeta}{\sum}w_i$ = 652.460, more than 26 times larger than any other GRBs in this set. The statistical significance of this detection can be estimated from the cumulative distribution of matched filter weights for a large sample of otherwise similar LAT fields. As shown in Figure 4, after including the GBM-triggered sample size of 74, the probability of such an event arising from random background photons is significantly less than 0.4\%. A more detailed discussion is provided in \S\ref{sec:LATObservations}
\begin{deluxetable}{llrrr}
 \tabcolsep 0.4mm
 \tablewidth{0pt}
 \tablecaption{List of 74 GBM triggered GRBs}
  \tablehead{\colhead{GRB} & \colhead{Trigger} & \colhead{RA} & \colhead{Dec} & \colhead{$S_{GBM}$ } \\
  \colhead{} & \colhead{} & \colhead{($^{\circ}$)} & \colhead{($^{\circ}$)} & \colhead{$\mu$erg/cm$^2$} }
\startdata
080805B & 080805496 & 322.70 &  47.90 &  -   \\
080822A & 080822647 &  63.60 &  25.80 &  -   \\
080824  & 080824909 & 122.40 &  -2.80 & 2.30 \\
080920  & 080920268 & 121.60 &   8.90 & 2.40 \\
081006A & 081006604 & 142.00 & -67.40 & 0.71 \\
081006B & 081006872 & 172.20 & -61.00 & 0.73 \\
081107  & 081107321 &  51.00 &  17.10 & 1.64 \\
081118B & 081118876 &  54.60 & -43.30 & 0.11 \\
081223  & 081223419 & 112.50 &  33.20 & 1.20 \\
081224  & 081224887 & 201.70 &  75.10 &  -   \\
081226C & 081226156 & 193.00 &  26.80 & 2.32 \\
081226B & 081226509 &  25.50 & -47.40 & 0.61 \\
081229  & 081229187 & 172.60 &  56.90 & 0.87 \\
081230B & 081230871 & 207.60 & -17.30 &  -   \\
090126B & 090126227 & 189.20 &  34.10 & 1.25 \\
090207  & 090207777 & 252.70 &  34.90 & 4.01 \\
090213  & 090213236 & 330.60 & -55.00 &  -   \\
090228B & 090228976 & 357.60 &  36.70 & 1.00 \\
090305B & 090305052 & 135.00 &  74.30 & 2.70 \\
090306C & 090306245 & 137.00 &  57.00 & 0.90 \\
090308B & 090308734 &  21.90 & -54.30 & 3.46 \\
090309B & 090309767 & 174.30 & -49.50 & 4.70 \\
090331  & 090331681 & 210.50 &   3.10 &  -   \\
090403  & 090403314 &  67.10 &  47.20 &  -   \\
090413  & 090413122 & 266.50 &  -9.20 &  -   \\
090429D & 090429753 & 125.21 &   6.20 & 1.60 \\
090519B & 090519462 & 105.90 & -56.70 & 1.40 \\
090529B & 090529310 & 231.20 &  32.20 & 0.34 \\
090617  & 090617208 &  78.89 &  15.65 & 0.47 \\
090623B & 090623913 &  41.70 &   1.80 &  -   \\
090625A & 090625234 &  20.29 &  -6.43 & 0.88 \\
090629  & 090629543 &   8.48 &  17.67 &  -   \\
090701  & 090701225 & 114.69 & -42.07 & 0.45 \\
090703  & 090703329 &   0.77 &   9.68 & 0.68 \\
090706  & 090706283 & 205.07 & -47.07 & 1.50 \\
090717B & 090717111 & 246.95 &  22.97 & 0.48 \\
090718A & 090718720 & 243.76 &  -6.68 &  -   \\
090726B & 090726218 & 240.45 &  36.75 &  -   \\
090807B & 090807832 & 326.90 &   7.23 & 1.02 \\
090815C & 090815946 & 251.26 &  52.93 &  -   \\
090819  & 090819607 &  49.08 & -67.12 &  -   \\
090820B & 090820509 & 318.26 & -18.58 & 1.16 \\
090826  & 090826068 & 140.62 &  -0.11 & 1.26 \\
090907B & 090907808 &  81.06 &  20.50 &  -   \\
090917  & 090917661 & 230.34 & -11.69 &  -   \\
091002  & 091002685 &  41.92 & -14.01 &  -   \\
091017B & 091017985 & 214.40 & -64.74 &  -   \\
091024B & 091024380 & 339.25 &  56.88 &  -   \\
091107  & 091107635 & 182.35 &  38.94 &  -   \\
091109C & 091109895 & 247.72 &  42.31 &  -   \\
091115  & 091115177 & 307.76 &  71.46 &  -   \\
091207A & 091207333 &  12.67 & -50.19 &  -   \\
091219  & 091219462 & 294.49 &  71.91 &  -   \\
091223A & 091223191 & 203.23 &  76.35 &  -   \\
091231B & 091231206 & 199.36 & -60.70 &  -   \\
100101B & 100101988 &  70.66 &  18.69 &  -   \\
100201A & 100201588 & 133.10 & -37.29 &  -   \\
100212B & 100212550 & 134.27 &  32.22 &  -   \\
100218A & 100218194 & 206.64 & -11.94 & 2.58 \\
100301A & 100301068 & 110.14 & -15.68 &  -   \\
100313B & 100313509 & 186.37 &  11.72 &  -   \\
100315A & 100315361 & 208.90 &  30.14 &  -   \\
100325B & 100325246 & 209.14 & -79.10 &  -   \\
100330B & 100330856 & 326.38 &  -6.97 &  -   \\
100401A & 100401297 & 281.85 & -27.83 & 2.39 \\
100417A & 100417166 & 261.31 &  50.38 &  -   \\
100427A & 100427356 &  89.17 &  -3.46 & 3.01 \\
100429A & 100429999 &  89.09 & -69.96 &  -   \\
100516A & 100516014 & 117.32 &  55.14 &  -   \\
100517B & 100517132 &  40.63 & -44.32 &  -   \\
100605A & 100605774 & 273.43 & -67.60 &  -   \\
100608A & 100608382 &  30.54 &  20.45 &  -   \\
100620A & 100620119 &  80.10 & -51.68 &  -   \\
100625B & 100625891 & 338.26 &  20.29 &  -   \\
\enddata
\end{deluxetable}

\begin{deluxetable}{crrrcr}
 \tabcolsep 0.4mm
 \tablewidth{0pt}
 \tablecaption{~ List of high energy photons for GRB 081006A}\label{tab:Tab_Highweight}
  \tablehead{\colhead{$i^a$} & \colhead{$t^b$} & \colhead{$\theta$ ($^\circ$)$^c$} & \colhead{$E$ (MeV) $^d$ } & \colhead{$c^e$} & \colhead{$w_i ^f$}}
\startdata
1  &  1.948  & 1.136 & 210.759 & 3 & 157.796  \\
2  &  6.384  & 1.253 & 130.083 & 3 & 138.073  \\
3  &  2.970  & 1.054 & 645.476 & 3 & 118.829  \\
4  &  13.251 & 0.809 & 787.895 & 3 & 59.160   \\
5  &  11.438 & 1.298 & 113.833 & 3 & 55.808   \\
6  &  2.048  & 0.372 & 283.650 & 2 & 51.225   \\
7  &  2.268  & 0.844 & 115.327 & 3 & 47.475   \\
8  &  6.806  & 1.348 & 148.296 & 2 & 27.328   \\
9  &  19.615 & 1.546 & 375.440 & 1 & 0.570    \\
10 &  3.076  & 1.762 & 567.849 & 1 & 0.354    \\
11 &  32.331 & 8.859 & 121.218 & 2 & 0.121    \\
12 &  26.478 & 0.132 & 773.543 & 3 & *0.094   \\
13 &  42.549 & 5.612 & 150.876 & 1 & 0.002    \\
\\
\multicolumn{6}{c}{GRB 081006A~~~~~${\zeta}$ = 0.99334~~~~~${\zeta}{\sum}w_i$ = 652.460}\\
\enddata
\tablenotetext{a}{photon ID number}
\tablenotetext{b}{time after the trigger}
\tablenotetext{c}{distance to the new estimated GRB location}
\tablenotetext{d}{energy of each photon}
\tablenotetext{e}{photon class}
\tablenotetext{f}{weight of each photon}
\tablenotetext{*}{indicates diminished $w_E$ for highest energy triplet cluster photon}
\end{deluxetable}

The triplet cluster finder associated with the matched filter technique estimates the location of GRB 081006A to be RA = 134.4$^{\circ}$ and Dec = -61.8$^{\circ}$ (J2000.0) with an uncertainty of $\sim$ 0.5$^\circ$. 13 high energy photons are spatially clustered in the nominal GRB 081006A direction. As shown in Table 2, 11 of these photons lie within a 2$^\circ$ radius. Figure 1 shows the LAT photon sky map of GRB 081006A. Our estimated direction (filled blue dot) is 6.6$^\circ$ away from the GBM determination (center of the map) which has a 1 $\sigma$ uncertainty of 4.5$^\circ$ with an additional systematic uncertainty of the order of 2.5$^\circ$ (van der Horst 2008), thus demonstrating consistency at the 1 $\sigma$ level. We note that GRB 081006A occurred $\sim$120s before the LAT ceased to take data as the satellite moved closer to the South Atlantic Anomaly (SAA). This must be accompanied by an increased photon background rate but over the whole LAT field before and after the GRB trigger we find no significant rate change that would create a spurious event that could mimic the signature of GRB 081006A.
\begin{figure}[!hbp]
\centering
   \includegraphics[width=.48\textwidth]{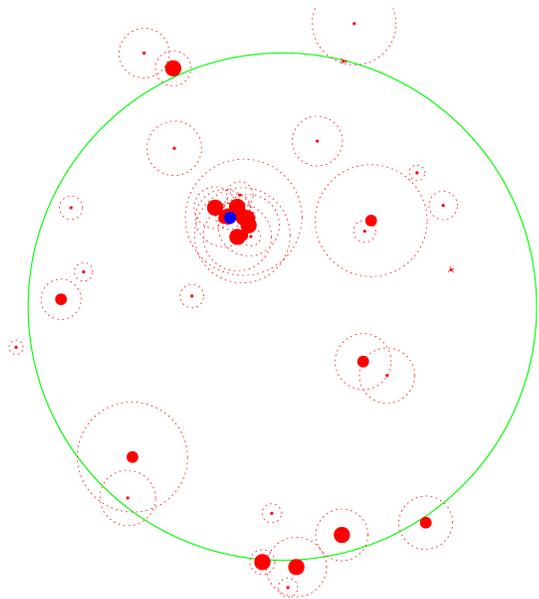}
   \caption{LAT high energy photons sky map for GRB 081006A. The diameter of each dot is proportional to its statistical weight. Thus, the largest diameters represent Event Class 3, etc. The dotted circles around each point indicate the $1-\sigma$ errors. The figure is centered on the nominal coordinates furnished by the GBM; the blue dot on the lower left shows the GRB coordinates computed by the cluster algorithm described in the text. The large green circle depicts the boundaries of the 16.0$^\circ$ radius cone that defines the fiducial boundaries for the cluster search. The plot axes are aligned so that North is up and East is to the right. \label{081006A_skymap}}
\end{figure}

\section{ $Fermi$ GBM and LAT observations}
\subsection{GBM observations}
At 14:29:34 UT on 6 October 2008, the $Fermi$ GBM triggered and located GRB 081006A (trigger 244996175/081006604; van der Horst 2008). Using the GBM trigger data, the on-ground calculated location was RA = 142.4$^{\circ}$, Dec = -67.4$^{\circ}$ (J2000) with an uncertainty of 4.5$^{\circ}$ (1 $\sigma$ containment, statistical only). The GRB $T_{90}$ is $\sim$7 s long (van der Horst 2008), clearly detected by NaI detector numbers 0 and 3. The BGO detector located on the same side, B0, also detected the increased flux.

Figure 2 shows the light curves obtained by the GBM and LAT. The upper plot is the background-subtracted light curve for the two NaI detectors, N0 + N3, in the 8 keV to 800 keV energy range. The counts were binned in 0.5 s intervals to provide a good visual representation of the intensity behavior. The middle plot is the background-subtracted light curve for the BGO (B0) detector in 280 keV to 4 MeV energy range with the same time binning. The GBM data shows a single peak concentrated in the first 2 s after the trigger. The lower plot is the light curve for the LAT detector above 100 MeV within an angle of 12$^{\circ}$ with respect
to the nominal GBM direction described above.

\subsection{LAT Observations}
\label{sec:LATObservations}
The boresight angle of GRB 081006A with respect to the LAT was 16$^{\circ}$ at the time of the trigger. The LAT did not slew to the burst direction but remained in normal observation mode, taking data for $\sim$120 s before entering the SAA. GRB 081006A was not reported by the LAT team but we have shown that it is unambiguously identified by the matched filter technique. With at least 13 high energy photons detected, it is bright enought to allow a likelihood analysis using the standard Science Tools software package\footnote{http://fermi.gsfc.nasa.gov/ssc/data/analysis/} provided by the FSSC.

The $gtlike$ tool, part of the standard Science Tools software package, generates the Test Statistic TS = 2$\Delta$log(likelihood) to compare models with and without an assumed source. The TS value associated with each source is a measure of the source significance or equivalently the probability that such an excess can be obtained from background fluctuations alone. A TS value of 25 corresponds to a 4.6$\sigma$ significance, approximately the square root of TS (Abdo et al. 2010a).

The first step to perform a likelihood analysis with the $gtlike$ tool is the selection of the LAT events belonging to the TRANSIENT class, re-constructed under current the Instrument Response Functions (IRF) : Pass6$\_$V3.
Following FSSC suggestions, the selected energy range spanned from 100 MeV to 100 GeV. The data below 100 MeV causes large uncertainties due to rapidly changing effective area with energy as well as ambiguity in the instrument response. We also set the maximum zenith angle value of 105$^{\circ}$ as suggested in the $Fermi$ data processing ``Cicerone\footnote{http://fermi.gsfc.nasa.gov/ssc/data/analysis/documentation/Cicerone/}". Finally, the time range for GRB data was set to be 0 - 50 s after the burst trigger time (T0), almost the same as described in Akerlof et al. (A10, A11).
\clearpage
\begin{figure}[!hbp]
\centering
   \includegraphics[width=.70\textwidth,angle=90]{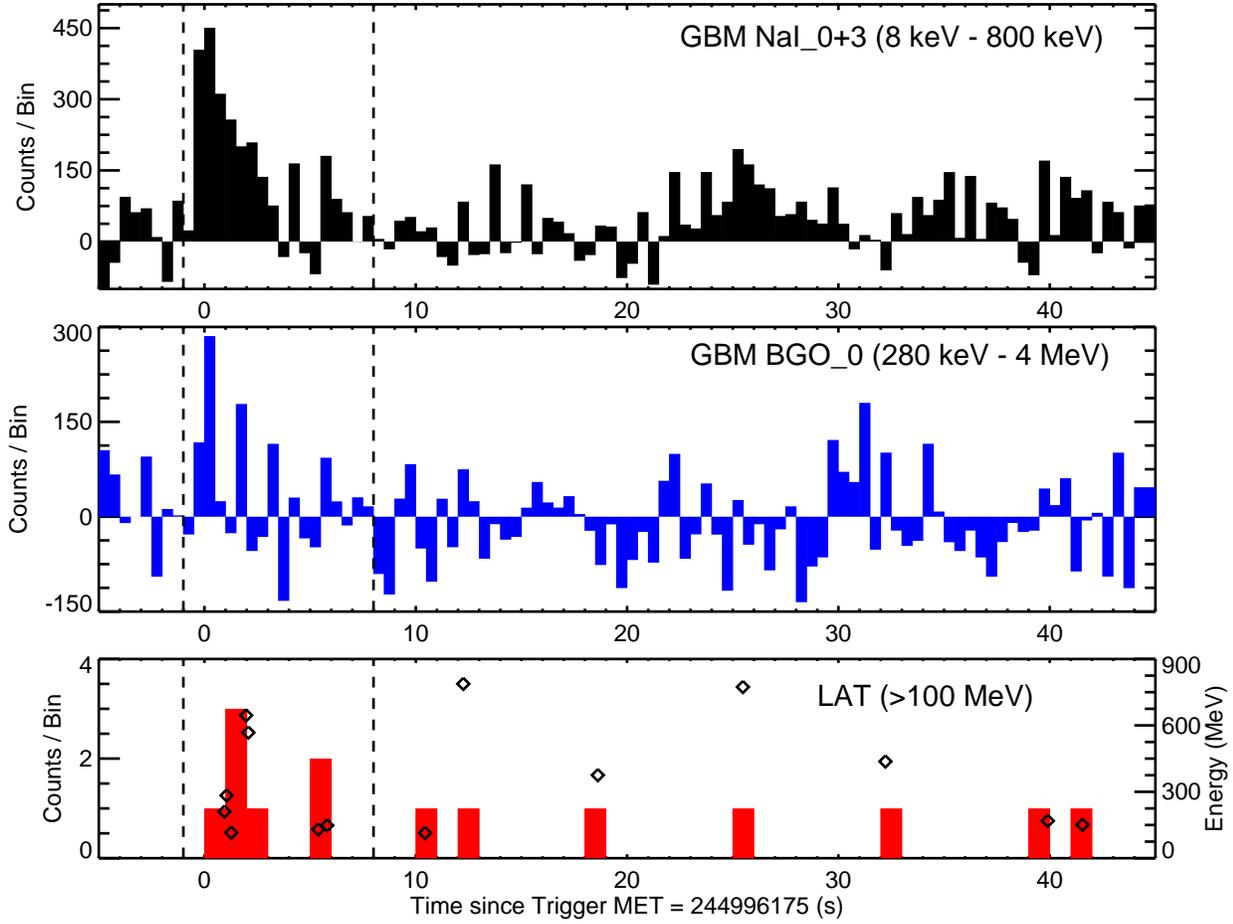}
   \caption{GBM and LAT light curves for GRB 081006A, in the order of increasing energy from top to bottom. The upper plot is the background subtracted sum of counts in the two NaI detectors, (N0 + N3), in the 8 keV to 800 keV energy range with a 0.5 s bin size. The middle plot is the background subtracted counts in the BGO (B0) detector in the 280 keV to 4 MeV energy range with a 0.5 s bin size. The lower plot shows the number of LAT events which passed the TRANSIENT event selection above 100 MeV,
   bin step is 1 s. The open diamonds represent the energy for each photon.} 
\end{figure}
\clearpage

The localization of GRB 081006A was also obtained from the TS map. The best fit position from the $gtfindsrc$ procedure is RA = 135.53$^{\circ}$ and DEC = -61.65$^{\circ}$ (J2000). Figure 3 shows the error contours around the fitted position with 68\%, 90\% and 99\% statistical error radii of 0.4$^{\circ}$, 0.6$^{\circ}$ and 0.9$^{\circ}$, respectively. This position is 0.55$^{\circ}$ from the location derived from the matched filter technique. Thus the locations derived from the two methods are consistent to within 1 $\sigma$ error.
\begin{figure}[!hbp]
\centering
   \includegraphics[width=.44\textwidth]{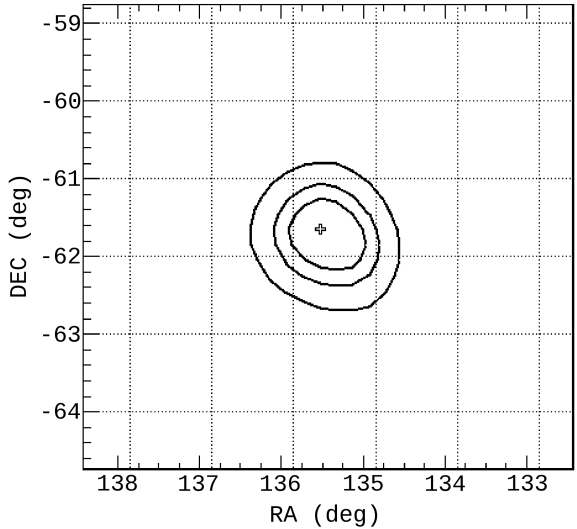}
   \caption{LAT localization for GRB 081006A with best fit position RA = 135.53$^{\circ}$ and DEC = -61.65$^{\circ}$ (J2000). The contours around the estimated position shows a 68\%, 90\% and 99\% statistical error radius of 0.4$^{\circ}$, 0.6$^{\circ}$ and 0.9$^{\circ}$, respectively.}
\end{figure}

The likelihood analysis indicates a detection with a TS value of 45 assuming an isotropic background model. Considering that GRB 081006A is close to the Galactic plane ($b \sim -0.9^{\circ}$), we must be careful that the high energy photon background does not have a local hot spot. However, a counts map with data taken three weeks prior to the burst demonstrated that the GRB position is safely $>$10$^{\circ}$ away from the bright region near the Galactic plane. We also estimated the count rate using the pre-burst data and re-scaled to the 50 s time range with TRANSIENT class events. Both these methods show that the background is not more than 3 photons in the 0 - 50 s time range. The TS detection value of 45 for GRB 081006A is close to the significance of GRB 080825C (Abdo et al. 2009d), GRB 081024B (Abdo et al. 2010b) and GRB 090217A (Ackermann et al. 2010).

In order to estimate the probability of such an occurrence by chance alone, we performed identical searches using the likelihood analysis method on 16088 random LAT fields that were selected similarly to the procedure described in A10. Considering that the likelihood analysis is extremely time consuming, about 15 minutes on a PC, we only applied the likelihood analysis to the 2192 random fields (out of the 16088) that contained more than three photons of class 2 or 3  within 16$^\circ$ radius region. If there are less than three class 2 or 3 photons are within the region, it is almost impossible to detect a reliable GRB signal. Figure 4 (upper panel) shows the cumulative distributions of the TS value for these 2192 random fields and the 27 (out of the 74) GRB fields that pass the same criteria. A K-S test indicates that the two distributions are consistent if we exclude the outlier, GRB 081006A. The random fields analysis shows that none have a TS value exceeding the candidate event, GRB 081006A.
The bottom panel of Figure 4 shows the cumulative distributions of the matched weight value (${\zeta}{\sum}w_i$) derived from the statistical technique described in A10.

In view of the heavy computational demands of the maximum likelihood method and the meager number of GRBs associated with high energy LAT photons, we decided to make some specific comparisons with the matched weight technique which is computationally at least a thousand times faster to compute. The distributions shown in Figure 4 indicate that the number of random fields that need to be examined to find a single event at the significance level of GRB 081006A is computationally daunting. Without embarking on such a brutal course the most conservative comparison of the statistical power of the two methods is obtained by noting that random probablity of obtaining a TS value greater than 45 is less than $0.0123 = 27/2192$ and the random probability of obtaining a matched weight value greater than 652 is less than $0.0043 = 74/17200$. Thus, the matched filter technique is at least three times more effective in rejecting background. A more realistic estimate can be obtained by assuming exponential decreases of the cumulative distributions with the relevant parameters. These estimated trends are shown by the cyan lines in the two graphs. This would suggest that the matched filter rejects background for similar events at a level at least 50 times better than the maximum likelihood approach. Thus the false positive probability for this GRB identification is of the order of $10^{-5}$. 
\begin{figure}[!hbp]
\centering
   \includegraphics[width=.45\textwidth]{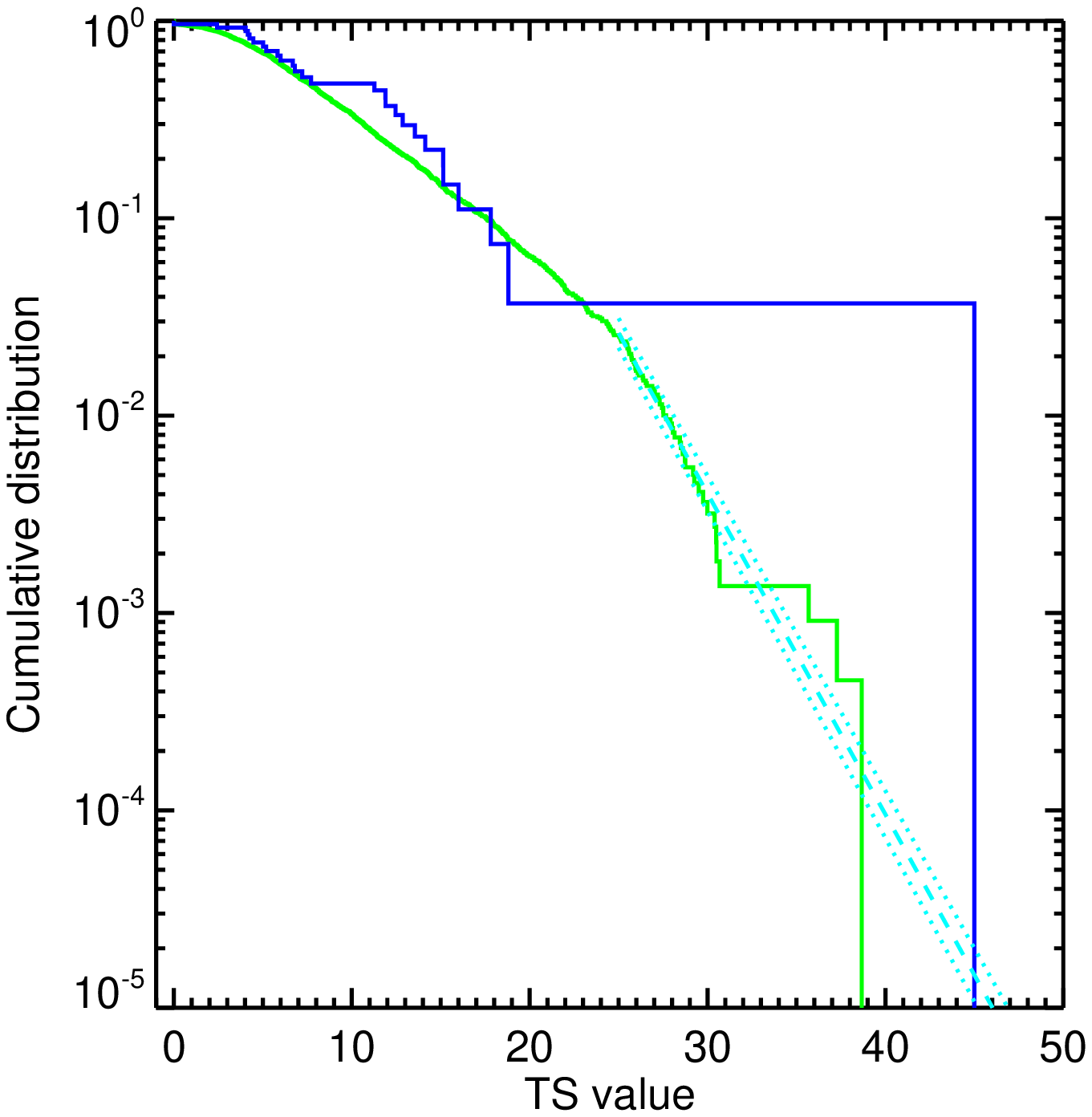}
   \includegraphics[width=.45\textwidth]{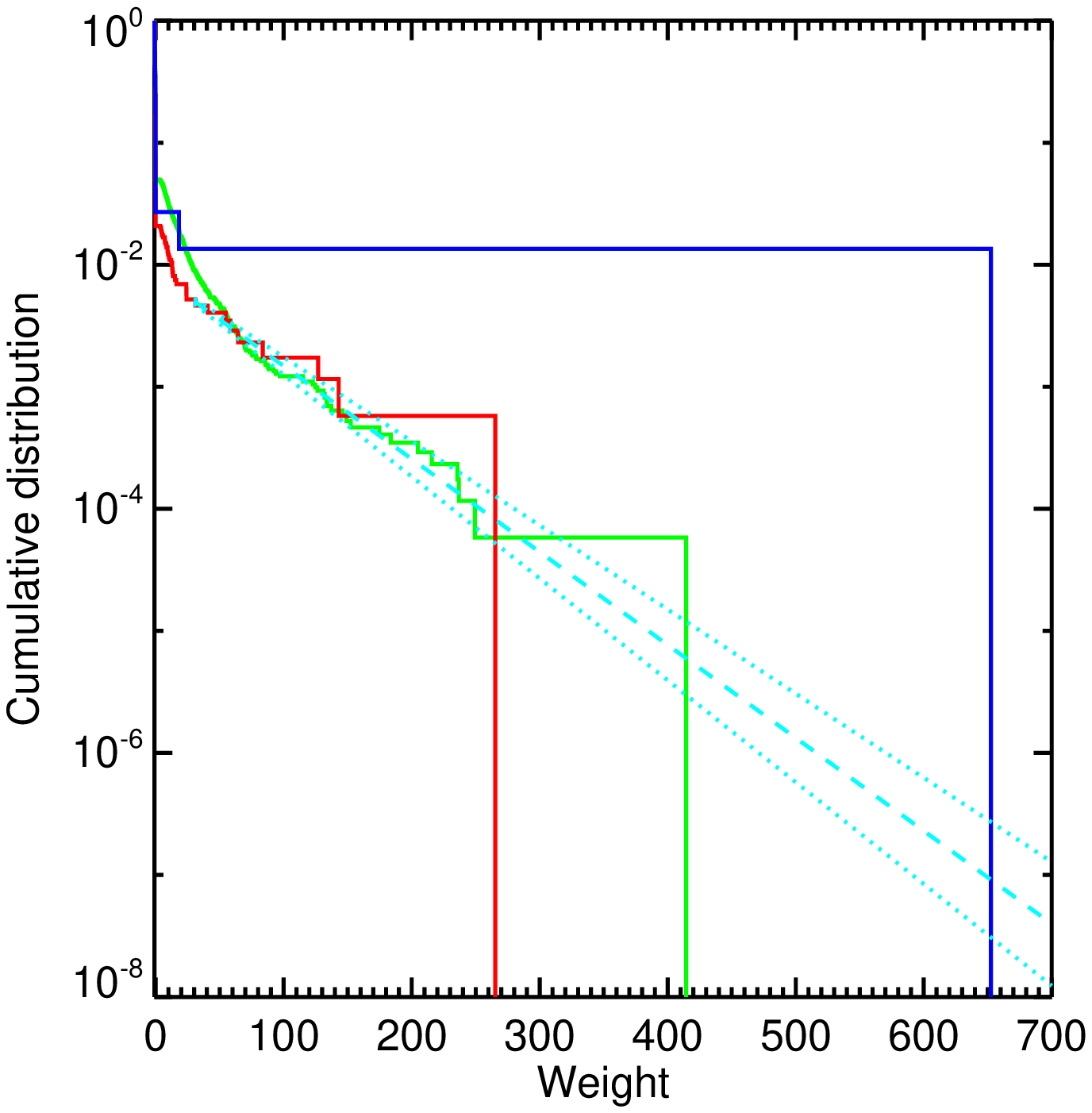}
   \caption{Upper panel : Complements of the cumulative distributions for TS values of 27 GBM GRBs (blue) and 2,192 similar fields obtained at random times (green). Bottom panel : The cumulative distributions for the matched weight values (${\zeta}{\sum}w_i$) of 74 GBM GRBs (blue), 1,731 random fields obtained nearly simultaneously with the GBM data (red), and 17,200 random fields obtained at random times (green). The dashed lines (cyan) in the two panels are extrapolations based on fits described in the text. (The dotted lines indicate estimated statistical
uncertainties.)
}
\end{figure}

With the localization obtained from the matched filter technique, the likelihood analysis confirmed the identification of faint LAT GRB 081006A. In several ways, these two methods are complimentary. The matched filter is computationally efficient for identifying weak signals with predetermined characteristics while the maximum likelihood method works best when the combinatorial background is relatively easy to estimate. The overall success of this approach is illustrated by GRB 081006A as well as the discoveries reported in previous papers.

\subsection{ Joint spectral fitting with GBM and LAT data}
Joint spectral fitting with the GBM and LAT data was performed for the time range, -1 s to 8 s, after the trigger. Due to the faintness of the burst, time-resolved spectral fitting is not appropriate. RMFIT (version 3.3) software was used for spectral fitting with binned GBM time-tagged event (TTE) data. The LAT data is re-binned to to match the format required by RMFIT.

Figure 5 shows the simultaneous fitting of the time-averaged count spectra from the GBM and LAT data.
The Band function model (Band et al. 1993) fit to the GBM and LAT data gives a reasonable good fit with amplitude A = 2.1$^{+0.6}_{-0.4}$ 10 $^{-3}$ (ph cm$^{-2}$ s$^{-1}$ keV$^{-1}$), E$_{p}$ = 817.0$^{+827}_{-340}$ (keV), $\alpha$ = 0.78$^{+0.35}_{-0.24}$ and $\beta$ = 2.28 $^{+0.10}_{-0.14}$ ($\chi^2$/dof = 464/379). The corresponding energy flux is 2.18$\pm 0.22$ 10$^{-7}$ erg cm$^{-2}$ s$^{-1}$ in the 10 keV to 1 MeV range and 6.06$\pm 0.61$ 10$^{-7}$ erg cm$^{-2}$ s$^{-1}$ in the 10 keV to 1 GeV range.
The E$_{p}$ of GRB 081006A is not well constrained, as already noticed by van der Horst (2008) but the Band function model fitting is clearly better than a single power-law model or a power-law with exponential cutoff model.
Given that some GRBs have more complex spectral behavior than described by the Band function (e.g. Zhang et al. 2011), we also tried a two component Band function plus power-law, and also the Band function plus thermal component but the fit was not improved.
\clearpage
\begin{figure}[!]
\centering
   \includegraphics[width=.57\textwidth,angle=90]{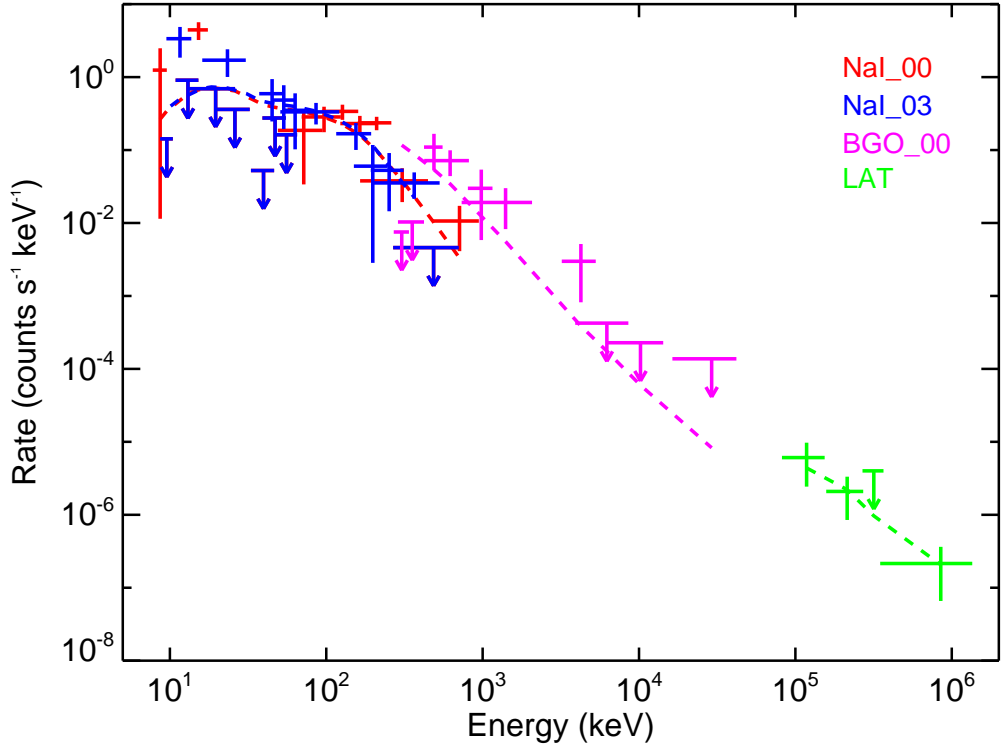}
   \caption{Spectral distribution of time-integrated (-1 s to 8 s) GRB 081006A photons obtained from GBM and LAT data. The spectrum is well fit by a Band function model spanning $\sim$5 decades of energy. The data from each instrument is indicated by color: $red$ = NAI 0, $blue$ = NAI 3, $pink$ = BGO 0 and $green$ = LAT. The dashed lines depict the fit obtained with the Band function model convolved with the four different instrument response functions.
}
\end{figure}
\clearpage

\section{Discussion}
GRB 081006A is detected with a fluence similar to three other GRBs, namely GRB 080825C, 081024B and 090217A. Among these events, GRB 081024B is a short duration burst while GRB 080825C and GRB 090217A are long duration. We thus compare the properties of GRB 081006A, also a long duration burst, with GRB 080825C and GRB 090217A.

The overall spectral features of GRB 081006A are quite similar to GRB 080825C and 090217A: they can be fitted by the Band function model over five decades energy range from the 8 KeV to 800 MeV. No cut-off feature in the LAT energy range is detected. This suggests that a single emission mechanism is responsible for the broadband emission of both GBM and LAT, e.g. synchrotron or jitter emission from the internal shock (Medvedev 2000).

The delayed onset of high energy LAT emission ($>$100 MeV), observed for many other LAT-detected GRBs ( e.g. GRB 080825C, Abdo et al. 2009d; GRB 080916C, Abdo et al. 2009b; GRB 090510, Abdo et al. 2009c; GRB 090902B Abdo et al. 2009a), also shows a hint in GRB 081006A. The initial peak of LAT high energy photons with GRB 081006A, though correlated with the GBM peak, is delayed for about 2 s after the low energy GBM trigger.
The delayed emission of high energy photons from GRB 081006A is not as pronounced as found for GRB 080825C. In the later case, the onset of LAT emission is occurs at about 3 s and the first LAT peak is coincident with the second GBM peak (Abdo et al. 2009d), but it is different from GRB 090217A for which there was no perceptible delay (Ackermann et al. 2010). 

Many GRBs ( e.g. GRB 080825C, Abdo et al. 2009d; GRB 080916C, Abdo et al. 2009b; GRB 090510, Abdo et al. 2009c; GRB 090902B Abdo et al. 2009a) also have long-lived high energy photon emission detected by the LAT even hours after the burst. The high energy photon emission of GRB 081006A clearly lasts longer ($\sim$ 40 s in LAT data) than the low energy emission (GBM T$_{90} \sim$ 7 s). The two highest energy photons, both with E $\sim$ 780 MeV, are detected at 13.25 s and 26.5 s after the trigger as shown in Table 2. This is also similar to GRB 080825C for which the LAT emission lasted slightly longer (up to T$_0$+35 s) than for the GBM (T$_{90}$ = 27 s). GRB 090217A does not show such longer higher energy emission.

\section{Summary}
We have demonstrated the association of GRB 081006A with high energy photon emission by applying the matched filter technique to the $Fermi$/LAT data. The false positive probability is definitely less than $4 \times 10^{-3}$ and
probably much smaller. This event is found to be correlated with at least 13 high energy photons detected by the LAT instrument. A maximum likelihood analysis reveals a similar confidence level with a TS value of 45. 
Comparing the temporal and spectral properties with the other two long duration GRBs with similar fluences, GRB 080825C and 090217A, we find GRB 081006A is closer to GRB 080825C. The delay and long emission duration for high energy photons are seen in GRB 081006A, similar to GRB 080825C, but not with GRB 090217A. These properties can be examined in more detail as the $Fermi$ mission continues to obtain a larger sample of GRBs, especially faint ones such as GRB 081006A, 080825C and 090217A.
As demonstrated here, the matched filter technique is considerably more sensitive than the maximum likelihood analysis to find these fainter events. In the particular case of GRB 081006A, the background rejection is approximately 50 times better, making it a far better search tool for these rare events. As we have shown with several original identifications of faint LAT GRBs, the two methods are best employed sequentially to first find the events and, secondly, to determine the event characteristics. In this unique period when both $Swift$ and $Fermi$ are operational, there is some urgency to explore the surprising dearth of faint LAT GRBs as fully as possible.

\vspace{0.1cm}

\acknowledgments
We thank Lin Lin at NAOC/UAH for valuable suggestions about the GBM data analysis. This research is supported by the NASA grant NNX08AV63G and the NSF grant PHY-0801007.


\begin{thebibliography}{}

\bibitem[{{Abdo} {et~al.}(2009a)}]{abdo09a} {Abdo}, A.~A., et al., 2009a, ApJ, 706, L138 

\bibitem[{{Abdo} {et~al.}(2009b)}]{abdo09b} {Abdo}, A.~A., et al., 2009b, Science, 323, 1688 

\bibitem[{{Abdo} {et~al.}(2009c)}]{abdo09c} {Abdo}, A.~A., et al., 2009c, Nature, 462, 331 

\bibitem[{{Abdo} {et~al.}(2009d)}]{abdo09d} {Abdo}, A.~A., et al., 2009d, ApJ, 707, 580 

\bibitem[{{Abdo} {et~al.}(2010)}]{abdo10} {Abdo}, A.~A., et al., 2010a, ApJS, 183, 46 

\bibitem[{{Abdo} {et~al.}(2010)}]{abdo10} {Abdo}, A.~A., et al., 2010b, arXiv:1002.3205 

\bibitem[{{Ackermann} {et~al.}(2010)}]{ackermann10} {Ackermann}, M., et al., 2010, ApJ, 717, L127 

\bibitem[{{Akerlof} {et~al.}(2010)}]{akerlof10} {Akerlof}, C., Zheng, W., Pandey, S.~B., McKay, T.~A., 2010, ApJ, 725, L15 (A10)

\bibitem[{{Akerlof} {et~al.}(2011)}]{akerlof11} {Akerlof}, C., Zheng, W., Pandey, S.~B., McKay, T.~A., 2011, ApJ, 726, 22 (A11)

\bibitem[{{Atwood} {et~al.}(2009)}]{atwood09} {Atwood}, W.~B., et~al., 2009, ApJ, 697, 1071

\bibitem[{{Band} {et~al.}(1993)}] {band93} {Band}, D., et~al. 1993, ApJ, 413, 281

\bibitem[{{Dingus } {}(1995)}] {dingus95} {Dingus}, B. L., 1995, Astrophys. Space Sci., 231, 187

\bibitem[{{Gehrels} {et~al.}(2004)}] {gehrels04} {Gehrels}, N., et al., 2004, ApJ, 611, 1005

\bibitem[{{Giuliani} {et~al.}(2008)}] {giuliani08} {Giuliani}, A., et al., 2008, A\&A, 491, 25

\bibitem[{{Granot} (2010)}] {granot07} {Granot}, J., 2007, arXiv:1003.2452 

\bibitem[{{Medvedev} (2000)}] {medvedev09} {Medvedev}, M.~V., 2000, ApJ, 540, 704 

\bibitem[{{Meegan} {et~al.}(2009)}] {meegan09} {Meegan}, C., et al., 2009, ApJ, 702, 791 

\bibitem[{{Paciesas} {et~al.}(2010)}] {paciesas10} {Paciesas}, W., et al., 2010, HEAD meeting \#11, \#10.03

\bibitem[{{van der Horst } (2008)}] {vander08} {van der Horst}, A.~J., 2008, GCN Circ., 8341 

\bibitem[{{Winkler} {et~al.}(2003)}] {winkler03} {Winkler}, C., et al., 2003, A\&A, 411, L1

\bibitem[{{Zhang}(2011)}] {zhang11a} {Zhang}, B. B., et al., 2011, ApJ, 730, 141 

\end{thebibliography}
\end{document}